\pdfoutput=1
\documentclass[a4paper,11pt]{article}
\usepackage{pos}
\usepackage{bbm}
\usepackage{xspace}
\usepackage{booktabs}
\usepackage{multirow}
\usepackage[export]{adjustbox}

\graphicspath{{plots/}}

\DeclareMathOperator{\tr}{tr}
\newcommand\AId{AI$^\dagger$\xspace}
\newcommand\AIId{AII$^\dagger$\xspace}
\newcommand{\eps}{\varepsilon}
\newcommand{\ev}[1]{\langle#1\rangle}
\newcommand\nrep{N_\text{rep}}
\renewcommand\phi\varphi
\newcommand{\RR}{\mathbbm{R}}
\newcommand{\ig}{}

\title{Complex spacing ratios of the non-Hermitian Dirac operator in universality classes \AId and \AIId}
\ShortTitle{Complex spacing ratios of the non-Hermitian Dirac operator}

\author[a]{Takuya Kanazawa}
\author*[b]{Tilo Wettig}

\affiliation[a]{Research and Development Group, Hitachi, Ltd., Kokubunji, Tokyo 185-8601, Japan}

\affiliation[b]{Department of Physics, University of Regensburg, 93040 Regensburg, Germany}

\emailAdd{tilo.wettig@ur.de}

\abstract{We consider non-Hermitian Dirac operators in QCD-like theories coupled to a chiral U(1) potential or an imaginary chiral chemical potential. We show that in the continuum they fall into the recently discovered universality classes \AId or \AIId of random matrix theory if the fermions transform in pseudoreal or real representations of the gauge group, respectively. For staggered fermions on the lattice this correspondence is reversed. We verify our predictions by computing spacing ratios of complex eigenvalues, whose distribution is universal without the need for unfolding.}

\FullConference{%
 The 38th International Symposium on Lattice Field Theory, LATTICE2021
  26th-30th July, 2021
  Zoom/Gather@Massachusetts Institute of Technology
}

\begin{document}
\maketitle

\section{Introduction}

Random matrix theory (RMT) is known to describe so-called universal features of eigenvalue spectra. Universal in this context means that these features are determined by global symmetries and independent of the details of the dynamics. Universal features have turned out to be useful in many applications in physics and beyond. In (lattice) QCD they can be used, e.g., to determine low-energy constants or to derive spectral sum rules \cite{Verbaarschot:2000dy}. In the last few years, interest in non-Hermitian systems has increased, and we shall see that these new developments are also relevant for (lattice) QCD.

Physical systems fall into distinct \emph{symmetry classes}. In the Hermitian case, there are 10 symmetry classes \cite{Altland:1997zz}, while in the non-Hermitian case there are 38 symmetry classes \cite{Kawabata:2018gjv,Zhou_2019}. The difference between these symmetry classes lies in certain anti-unitary symmetries and in the behavior of the spectrum near the origin. However, if we concentrate on short-range correlations in the bulk of the spectrum, different symmetry classes can yield the same results, and it turns out that both in the Hermitian and in the non-Hermitian case the bulk spectral correlations are described by only 3 distinct \emph{universality classes}. In Hermitian RMT, these are the well-known Wigner-Dyson ensembles, i.e., the Gaussian Unitary, Orthogonal and Symplectic Ensembles (GOE, GUE, GSE), while in non-Hermitian RMT the universality classes are called Ginibre, \AId and \AIId \cite{Ginibre:1965zz,jaiswal2019universality,Hamazaki_2020}. The latter differ in their transposition symmetries: the Ginibre class does not possess such a symmetry, in \AId the matrices are complex symmetric ($X^T=X$),  and in \AIId the matrices satisfy $X^T=\sigma_2X\sigma_2$.

The Ginibre class has been known for a long time, and it was shown that the spectral correlations of the lattice QCD Dirac operator at nonzero chemical potential follow the Ginibre predictions \cite{Markum:1999yr}. In this contribution, we will show that the recently discovered classes \AId and \AIId are also realized in the spectrum of the Dirac operator for QCD-like theories in the continuum and on the lattice \cite{Kanazawa:2021zmv}.

\section{Non-Hermitian Dirac operators: continuum and lattice symmetries}
\label{sec:symm}

We take the continuum Euclidean Dirac operator and couple it to a chiral U(1) gauge field $B$,
\begin{align}
  \label{eq:Dcont}
  D = \gamma_\nu(\partial_\nu - i A^a_\nu \tau_a - i \gamma_5 B_\nu)\,.
\end{align}
Here, the $\gamma_\nu$ are the Euclidean Dirac matrices with $\gamma_5=\gamma_1\gamma_2\gamma_3\gamma_4$,  $A_\nu^a$ is the usual gauge field, and the $\tau_a$ are the generators of the gauge group. If instead of a chiral U(1) field we add $B_\nu=-i\mu_5\delta_{\nu4}$ with a chiral chemical potential $\mu_5$, the last term in \eqref{eq:Dcont} becomes $\mu_5\gamma_5\gamma_4$ and represents a chirality imbalance (for real $\mu_5$) or a source term for a spatially inhomogeneous chiral condensate (for imaginary $\mu_5$) \cite{Kanazawa:2021zmv}. For $B=0$ (or $\mu_5\in\RR$) the eigenvalues of $D$ are purely imaginary, but for $B\ne0$ (or $\mu_5\notin\RR$) they are generically complex.  We now consider gauge groups with pseudoreal and real representations, such as SU(2) for which the fundamental representation is pseudoreal and the adjoint representation is real, respectively. In the following, $K$ denotes the operator of complex conjugation and $C=i \gamma_4 \gamma_2$ the charge-conjugation operator. For pseudoreal representations, the Dirac operator without the $B$ field has the antiunitary symmetry $[iD, C \tau_2 K] = 0$. This symmetry is broken by $B$, but $D$ retains the transposition symmetry $D^T = C \tau_2 D C \tau_2$. In this case a basis can be chosen in which $D^T=D$. For real representations, the Dirac operator without $B$ has the antiunitary symmetry $[iD,CK]=0$. Again this is broken by $B$, but the transposition symmetry $D^T=CDC$ remains. In this case a basis can be chosen in which $D^T=\sigma_2D\sigma_2$, where $\sigma_2$ is the second Pauli matrix acting on the Dirac indices. In Table~\ref{tab:cont} we summarize the symmetry properties of the continuum Dirac operator and the form of the corresponding random matrix.
\begin{table}
  \centering
  \begin{tabular}{@{}c@{\hspace*{7mm}}c@{\hspace*{7mm}}c@{\hspace*{7mm}}
    c@{\hspace*{7mm}}c@{}}
    \toprule
    Representation & Symmetry of $D$ & Matrix form & Matrix elements & Class\\
    \midrule
    pseudoreal & $D^T=D$ & $\begin{pmatrix}0&V\\V^T&0\end{pmatrix}$ & complex
    & \AId\\[5mm]
    real & $D^T=\Sigma_2D\Sigma_2$ & $\begin{pmatrix}0&V\\\sigma_2V^T\sigma_2&0\end{pmatrix}$
    & \multirow[t]{2}[6]{*}{\parbox{20mm}{%
      \begin{center}complex\\quaternion\end{center}}} & \AIId\\
    \bottomrule
  \end{tabular}
  \caption{Symmetries of the continuum Dirac operator with chiral U(1) field in a suitable basis and corresponding random matrix ensembles. Here, $\Sigma_2=\sigma_2\oplus\sigma_2$. The block structure is due to the fact that we always have chiral symmetry, $\{D,\gamma_5\}=0$. For \AIId, every eigenvalue is twofold degenerate (Kramers degeneracy).}
  \label{tab:cont}
\end{table}

We now turn to the lattice and consider the staggered Dirac operator,\footnote{We use the same symbol $D$ for the continuum and lattice Dirac operator.} which has the remnant chiral symmetry $\{D,\eps\}=0$ with $\eps_{xy}=\eps(x)\delta_{xy}$ and $\eps(x)=(-1)^{x_1+x_2+x_3+x_4}$. Coupling the staggered Dirac operator to a chiral U(1) field $\theta_\mu(x)=\exp(i\eps(x)\phi_\mu(x))$ with real $\phi_\mu(x)$ we have
\begin{align}
  \label{eq:stagg}
  D(\theta)_{xy} = \frac{1}{2} \sum_{\mu=1}^{4}\eta_\mu(x)
  \left[U_\mu(x)\theta_\mu(x)\delta_{x+\mu,y} - U_\mu(y)^\dagger\theta_\mu(y)\delta_{x,y+\mu}\right]
\end{align}
with the usual link variables $U_\mu(x)$ and the staggered phases $\eta_\mu(x)$.
We consider SU(2) gauge fields in the fundamental and adjoint representation, denoted by $U^F$ and $U^A$, respectively. They are related by
\begin{align}
  U_\mu^A(x)_{ab}=\frac12\tr(\tau_aU_\mu^F(x)\tau_bU_\mu^F(x)^\dagger)\,.
\end{align}
In the presence of the chiral U(1) field, the transposition symmetries are now given by $D^F(\theta)^T=-\tau_2D^F(\theta)\tau_2$, which corresponds to class \AIId, and $D^A(\theta)^T=-D^A(\theta)$, which corresponds to class \AId, respectively.\footnote{The minus signs in these two relations lead to an additional factor of $-1$ in one of the off-diagonal blocks in Table~\ref{tab:cont}. This gives a relative factor of $i$ in the eigenvalues and leaves the bulk spectral correlations unchanged.} Hence the staggered symmetries are reversed in these two cases compared to the continuum, just as in the Hermitian case \cite{Halasz:1995vd}.

Instead of the chiral U(1) field we can also introduce a chiral chemical potential $\mu_5$. Following \cite{Braguta:2015zta} but with slightly different notation we have
\begin{align}
  D(\mu_5)_{xy}=D(\theta=1)_{xy}
  +\frac12\mu_5s(x)\left[\bar U_\delta(x)\delta_{x+\delta,y}+\bar U_\delta(y)^\dagger\delta_{x,y+\delta}\right],
\end{align}
where the first term on the RHS is the usual staggered operator, i.e., Eq.~\eqref{eq:stagg} with $\theta_\mu(x)=1$, and
\begin{align}                 
  s(x)=(-1)^{x_2}\,, \quad \delta=(1,1,1,0)\,,\quad
  \bar U_\delta(x)=\frac16\sum_{i,j,k=\text{perm}(1,2,3)}U_i(x)U_j(x+\hat i)U_k(x+\hat i+\hat j)\,.
\end{align}
This operator has the same symmetries as $D(\theta)$. In the continuum limit, the $\mu_5$ term gives $\mu_5\gamma_5\gamma_4$ as required  \cite{Braguta:2015zta}. For $\mu_5\notin\RR$ the eigenvalues move into the complex plane, and in our simulations we use $\mu_5\in i\RR$.

Exact spectral sum rules are a useful check for the correct computation of eigenvalues. 
For the standard massless staggered Dirac operator it is well known that
\begin{align}
  \tr D^2=\sum_n\lambda_n^2=-2\nrep V\,,
\end{align}
where $V$ is the lattice volume and $\nrep$ is the dimension of the repesentation of the gauge field in which the fermions transform, e.g., $\nrep=2$ for SU(2) fundamental and $\nrep=3$ for SU(2) adjoint. The generalization to a chiral U(1) field and a chiral chemical potential reads 
\begin{align*}
  \tr D^2=\sum_n\lambda_n^2=-V\left[2\nrep\ev{\theta_\mu^2(x)}_{x\mu}
  +\frac12\mu_5^2\bigl\langle\tr\bar U_\delta(x)\bar U_\delta(x)^\dagger\bigr\rangle_x\right],
\end{align*}
where $\ev{\cdots}_{x\mu}$ and $\ev{\cdots}_x$ denote averages over links and sites, respectively. Note that $\bar U_\delta(x)$ is not unitary. All of our numerical results were checked against this sumrule. A mass term can be added trivially \cite{Kanazawa:2021zmv}.

\section{Complex spacing ratios}

To test our predictions for the universality classes of the staggered Dirac operator with chiral U(1) field or chiral chemical potential we compute all eigenvalues of the operator in the complex plane. A well-known universal quantity is the nearest-neighbor spacing distribution $P(s)$, but to construct this quantity the eigenvalues would need to be unfolded first. Unfolding in the complex plane is a difficult problem \cite{Markum:1999yr,Akemann_2019}, and recently another universal quantity was proposed \cite{SRP2020} that avoids this problem, i.e., the distribution of the complex spacing ratios
\begin{align}
  \label{eq:nn}
  z_k=\frac{\lambda_k^\text{NN}-\lambda_k}{\lambda_k^\text{NNN}-\lambda_k}\,,
\end{align}
where $\lambda_k^\text{NN}$ and $\lambda_k^\text{NNN}$ are the nearest and next-to-nearest neighbor of $\lambda_k$ in the complex plane. Writing $z=re^{i\theta}$ we have $|r|\le1$ by construction.\footnote{The angle $\theta$ is not to be confused with the chiral U(1) field $\theta_\mu(x)$ in Sec.~\ref{sec:symm}.}
The distribution $P(z)$ is universal and described by RMT. Analytical results are not available yet, and therefore we generated the RMT predictions numerically from random matrices of dimension $N=4000$ with a Gaussian distribution of the matrix elements.
We also compute the marginal distributions $P(r)$ and $P(\theta)$ (obtained by integrating over $d\theta$ and $rdr$, respectively) as well as several moments of $P(z)$.

\section{Numerical results}

We performed lattice simulations with gauge group SU(2) and staggered fermions using the Grid/gpt framework \cite{Boyle:2015tjk,gpt}. We considered $2\cdot3=6$ cases. The factor of 2 refers to the two representations of SU(2) (fundamental and adjoint) in which the fermions transform. For each representation we included either a chiral U(1) field or a chiral chemical potential. The factor of 3 arises since we considered two different values of the latter ($\mu_5=i$ and $\mu_5=2i$ in lattice units).

Our simulations are quenched, which is sufficient since the presence of the fermion determinant does not change the universality class. The coupling constants were chosen to be $\beta_\text{SU(2)}=2.0$ and $\beta_\text{U(1)}=0.9$, corresponding to the confined phase. For the generation of the SU(2) and U(1) fields we employed the Creutz--Kennedy-Pendleton \cite{Creutz:1980zw,Kennedy:1985nu} and Hattori-Nakajima \cite{Hattori:1992qk} heatbath algorithms, respectively. Our lattice volume is $8^3\times16$, for which we can easily compute all eigenvalues. The number of configurations is given in Table~\ref{tab:conf}. For the nearest-neighbor search needed in Eq.~\eqref{eq:nn} we used a $k$-d tree algorithm.

\begin{table}
  \centering
  \begin{tabular}{@{}l@{\hspace{.05\columnwidth}}@{\hspace{.05\columnwidth}}
    c@{\hspace{.05\columnwidth}}c@{\hspace{.05\columnwidth}}c@{}}
    \toprule
    Representation & U(1) & $\mu_5=i$ & $\mu_5=2i$ \\
    \midrule
    SU(2) fundamental & 271 & 415 & 415 \\
    SU(2) adjoint & 267 & 267 & 266 \\
    \bottomrule
  \end{tabular}
  \caption{Number of configurations for the six cases we considered. The lattice volume is always $8^3\times16$.}
  \label{tab:conf}
\end{table}

\begin{figure}[b]
  \centering
  \renewcommand\ig[1]{\includegraphics[width=.25\textwidth]{#1}}
  \begin{tabular}{c@{\hspace{5mm}}c@{\hspace{5mm}}c}
    \ig{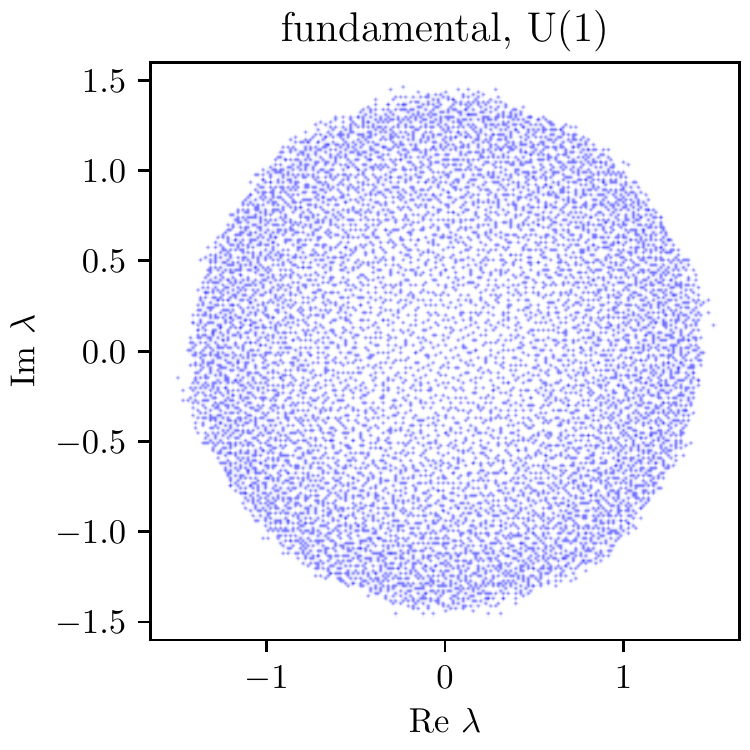} & \ig{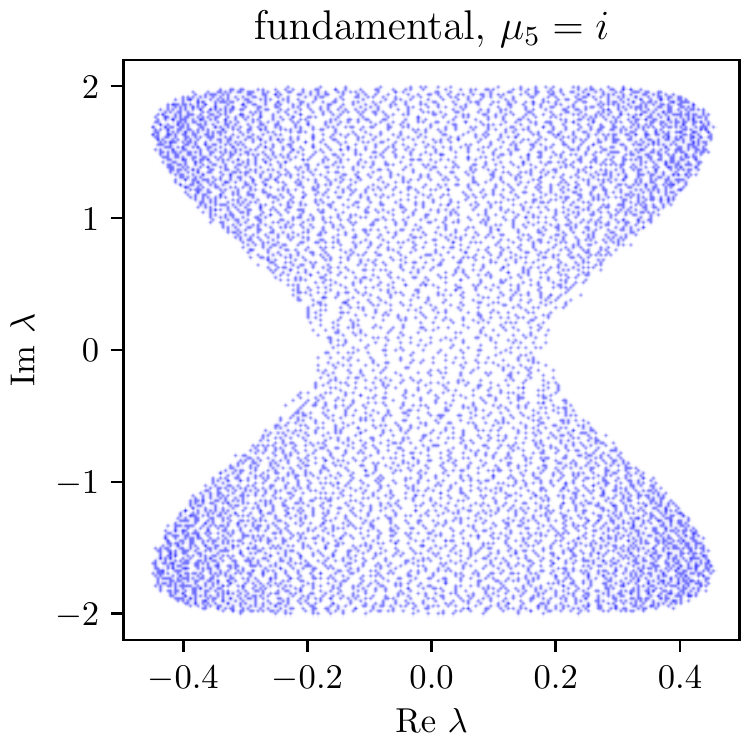} & \ig{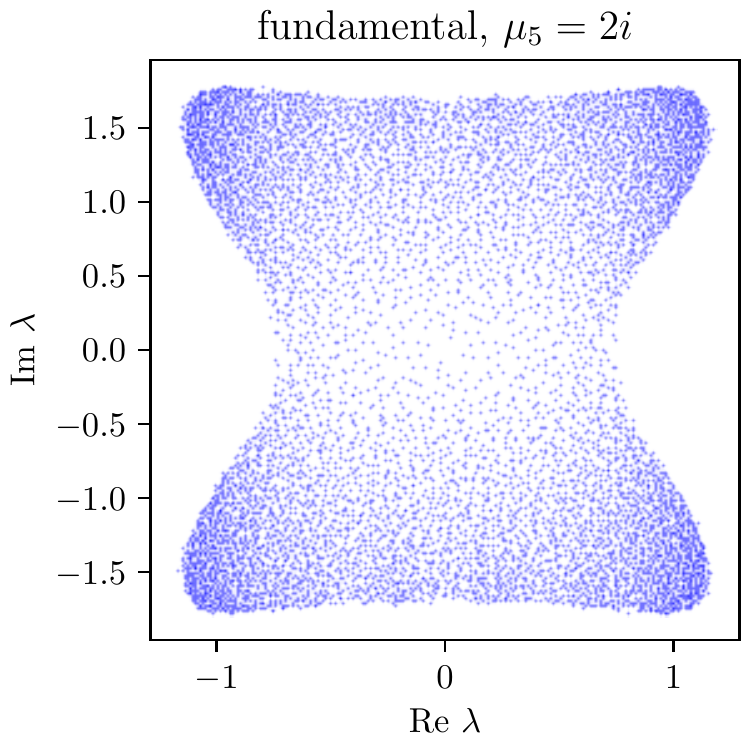}\\
    \ig{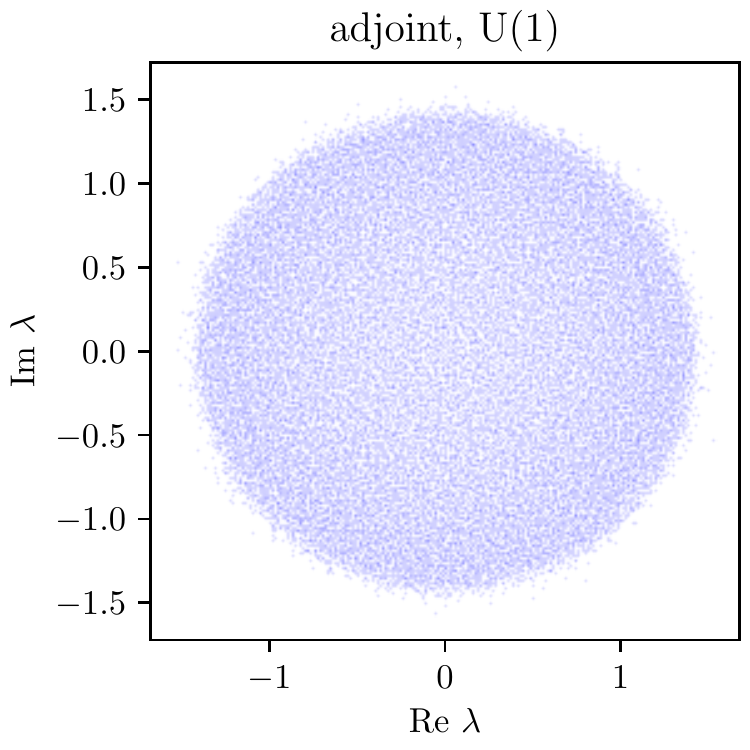} & \ig{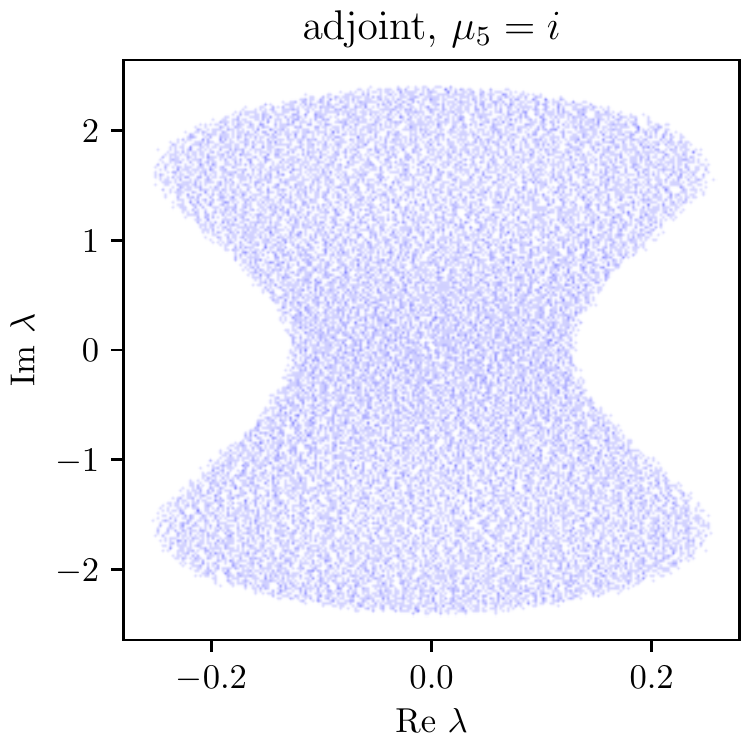} & \ig{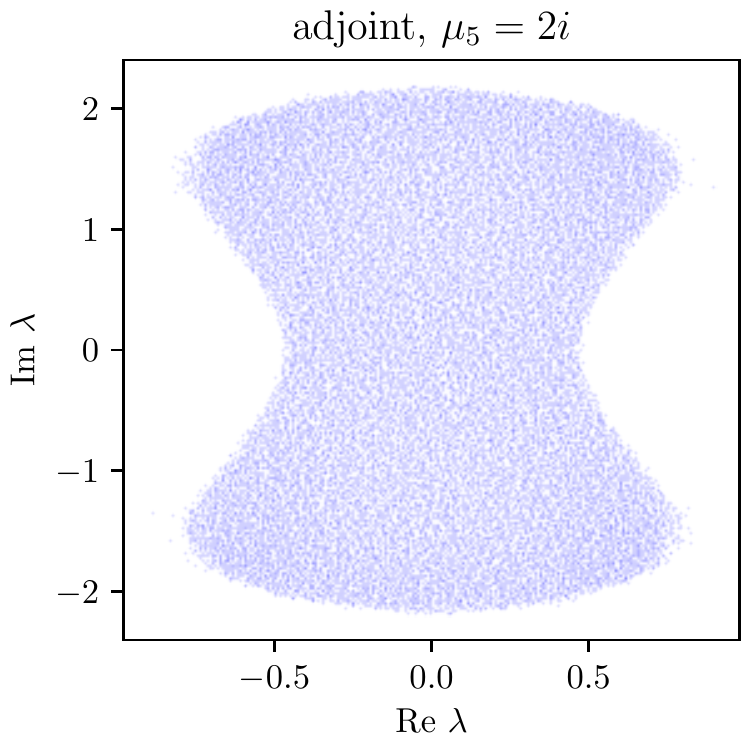}
  \end{tabular}
  \caption{Scatter plots of the complex Dirac eigenvalues for typical configurations. }
  \label{fig:scatter}
\end{figure}

In Figure~\ref{fig:scatter} we show typical scatter plots of the eigenvalues. As expected, the spectra are non-universal, i.e., the shape of the support of the spectrum and the eigenvalue density depend on the simulation parameters. The main purpose of this figure is to show that the density is smooth and the support is simply connected. Hence Eq.~\eqref{eq:nn} can be applied without the need to consider separate clusters of eigenvalues.

\begin{figure}
  \centering
  \renewcommand\ig[1]{\includegraphics[width=.25\textwidth,valign=c]{#1}}
  \begin{tabular}{l@{\hspace*{.05\textwidth}}ccc}
    RMT & \ig{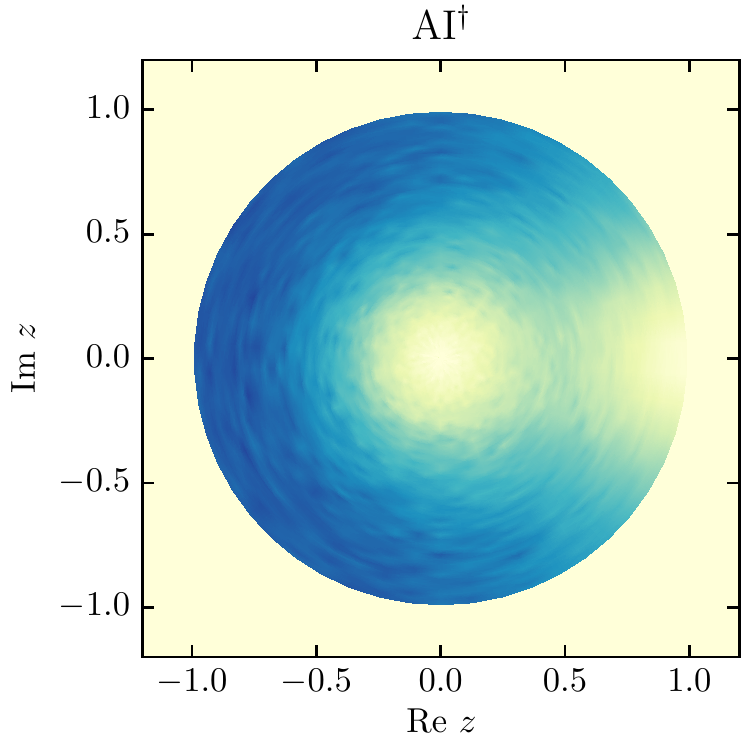} & \ig{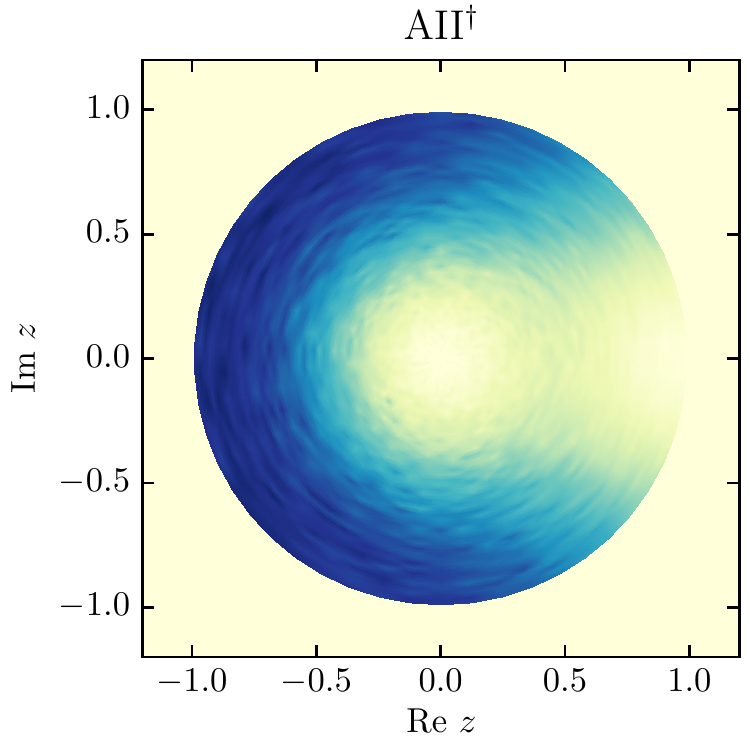} & \ig{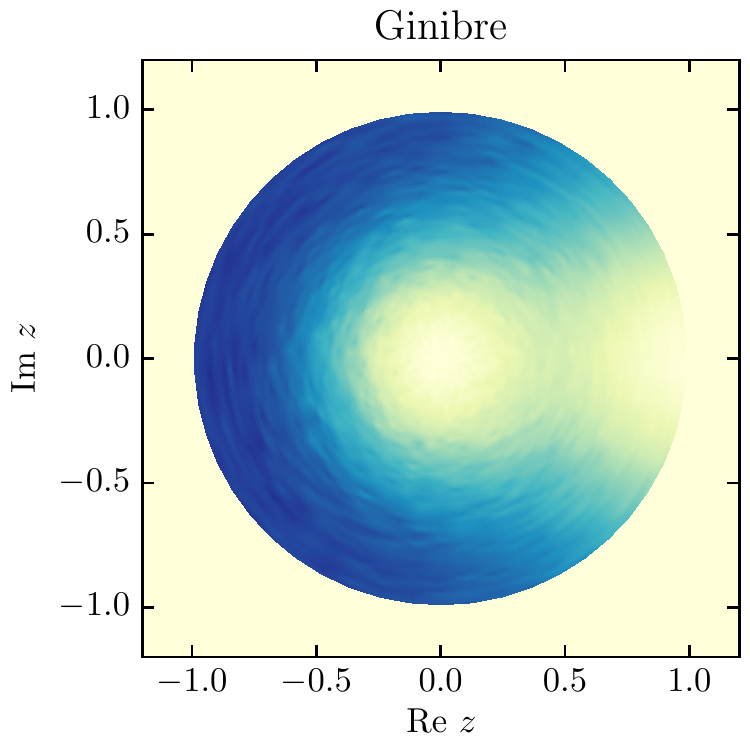}\\
    \parbox[c]{30pt}{SU(2)\\fund.} & \ig{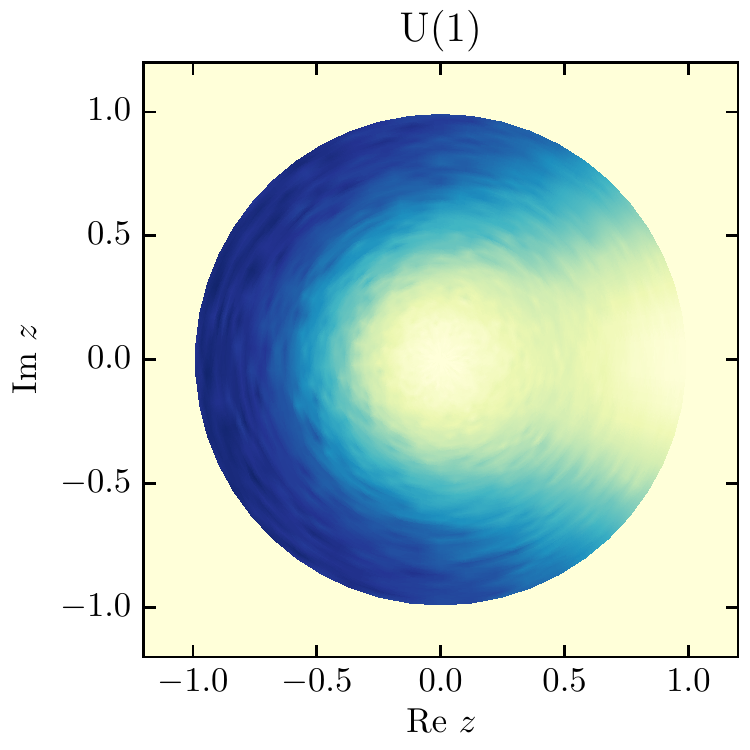} & \ig{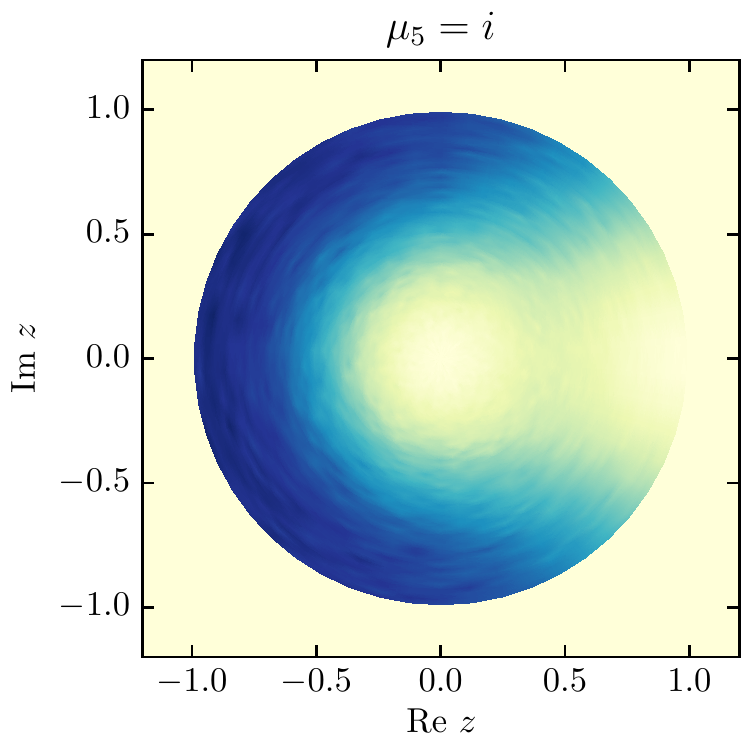} & \ig{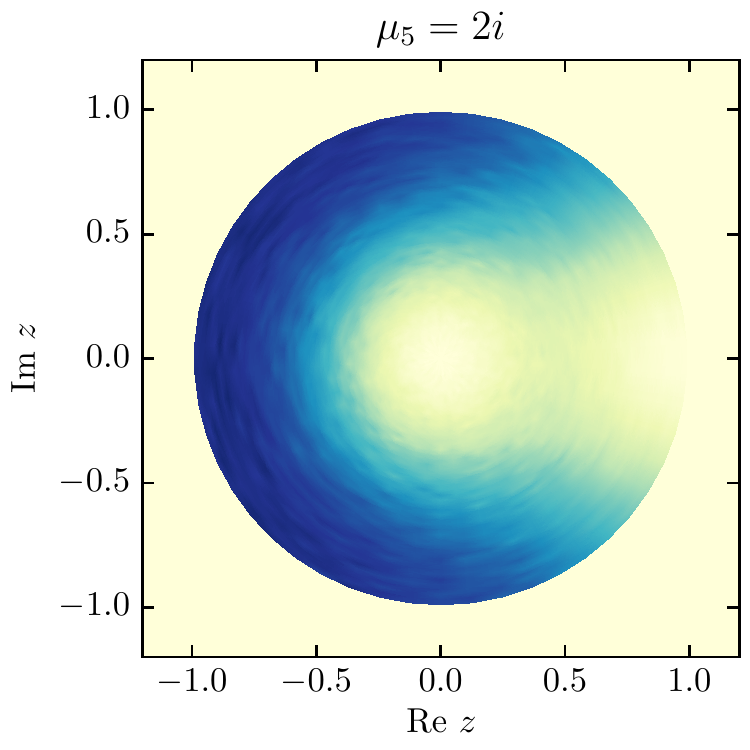}\\
    \parbox[c]{30pt}{SU(2)\\adjoint} & \ig{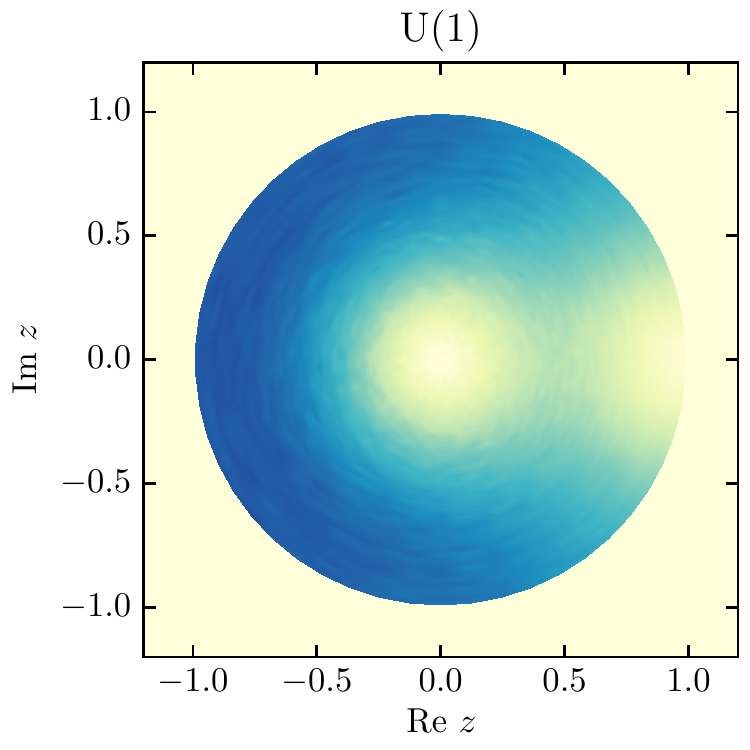} & \ig{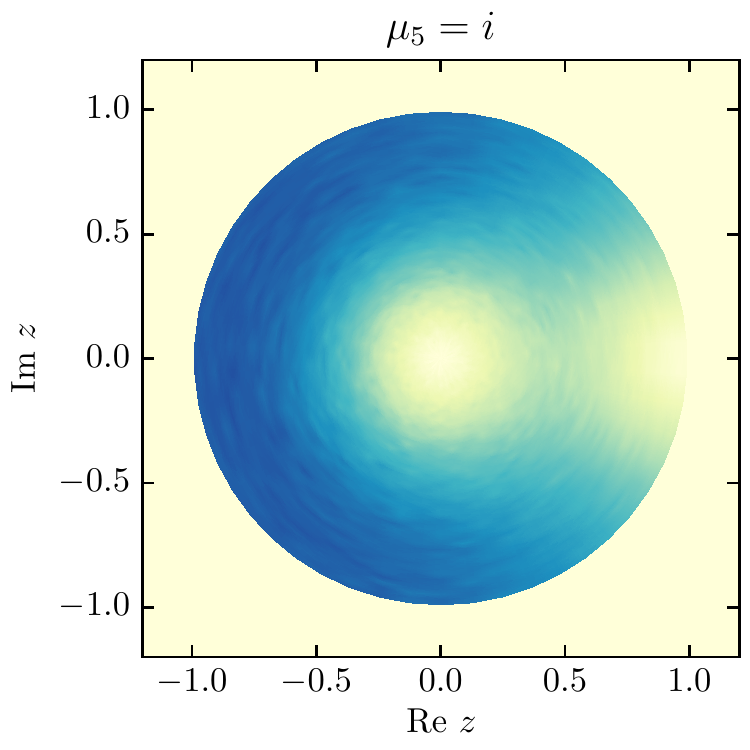} & \ig{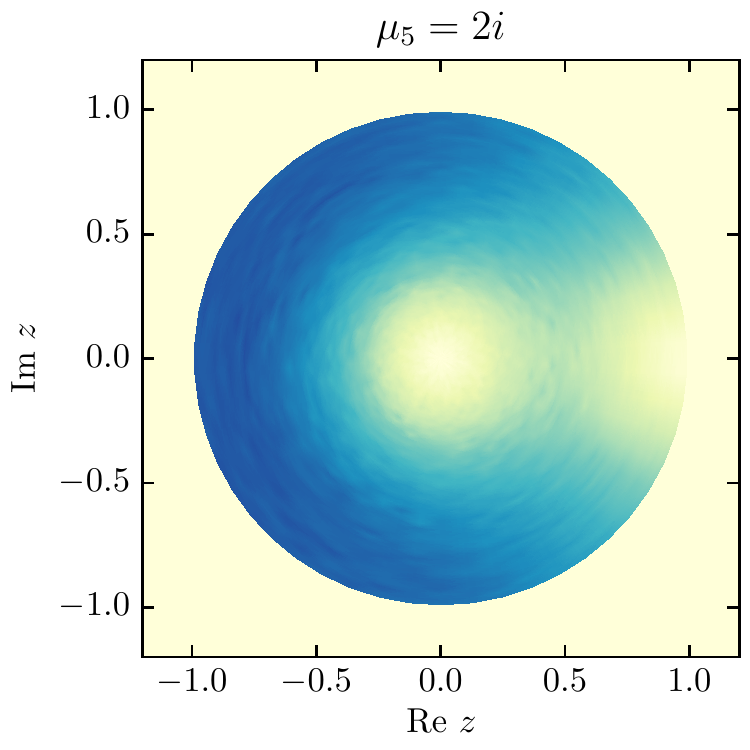}
    \end{tabular}      
    \caption{Probability distribution $P(z)$ of the complex spacing ratio $z$ in Eq.~\eqref{eq:nn}. The top row shows the three universality classes of RMT. The middle row shows lattice results for SU(2) in the fundamental representation. In all three cases $P(z)$ agrees with universality class \AIId. The bottom row shows the lattice results for the adjoint representation of SU(2), which agree with universality class \AId.}
  \label{fig:2d}
\end{figure}

\begin{figure}
  \centering
  \renewcommand\ig[1]{\includegraphics[width=.28\textwidth,valign=c]{#1}}
  \parbox[c]{.595\textwidth}{
    \begin{tabular}{c@{\hspace*{.2em}}c}
      \ig{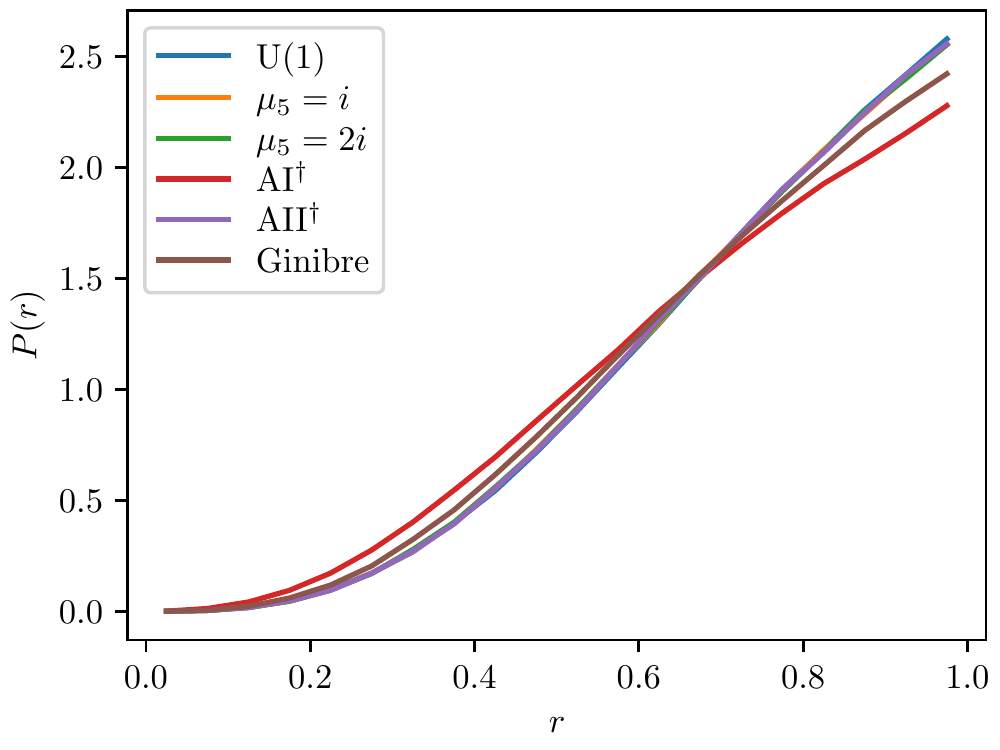} & \ig{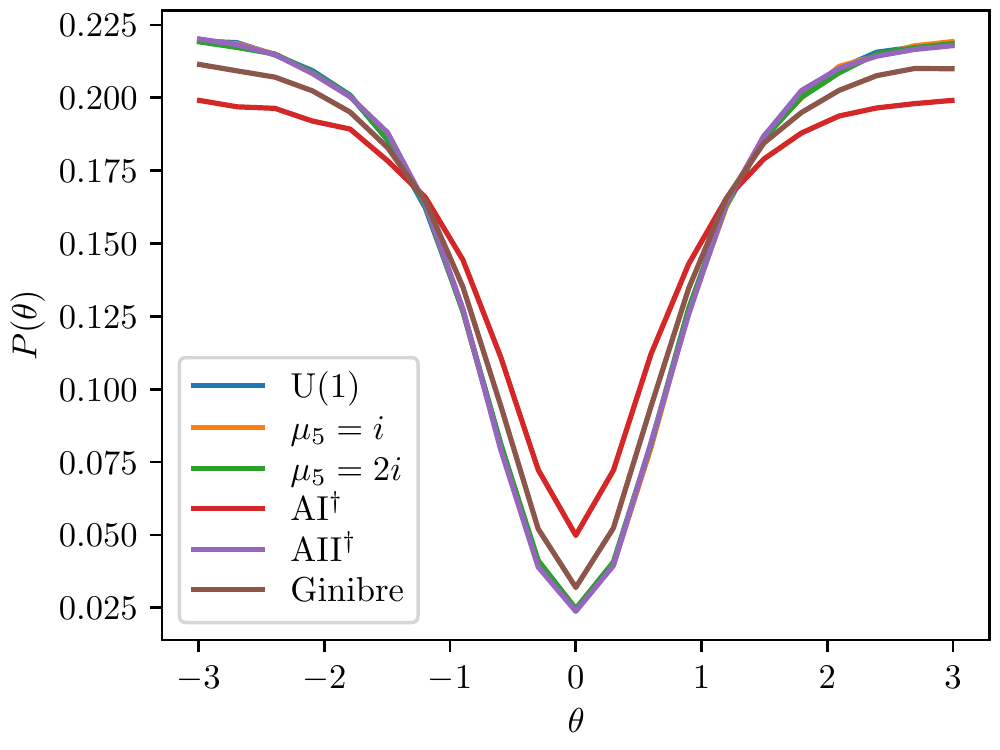} \\
      \ig{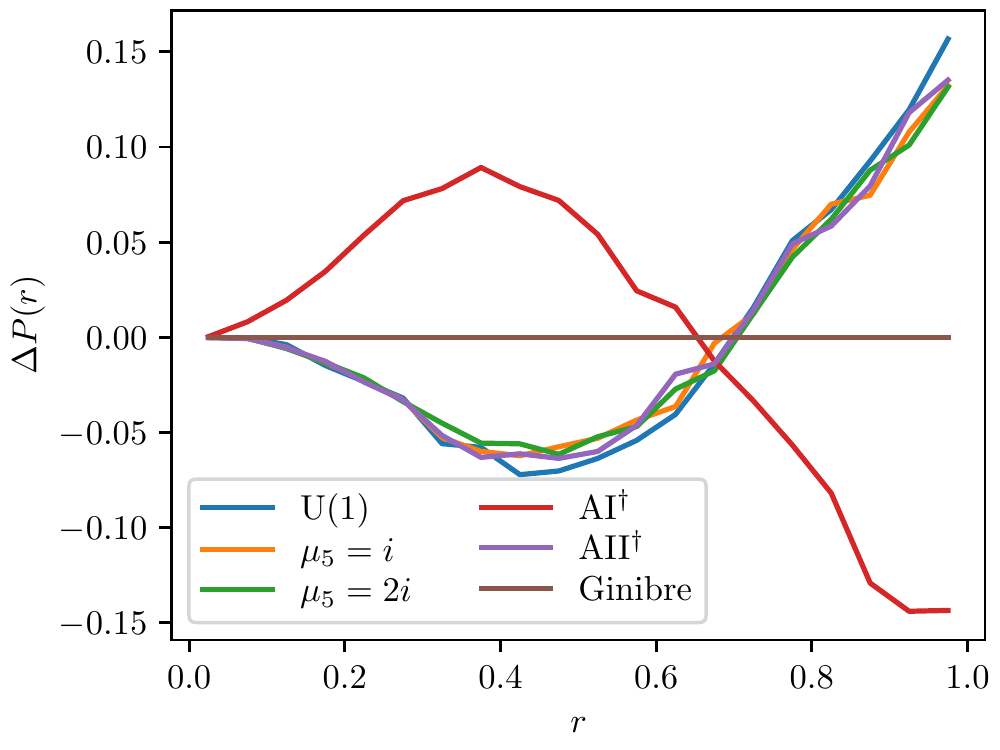} & \ig{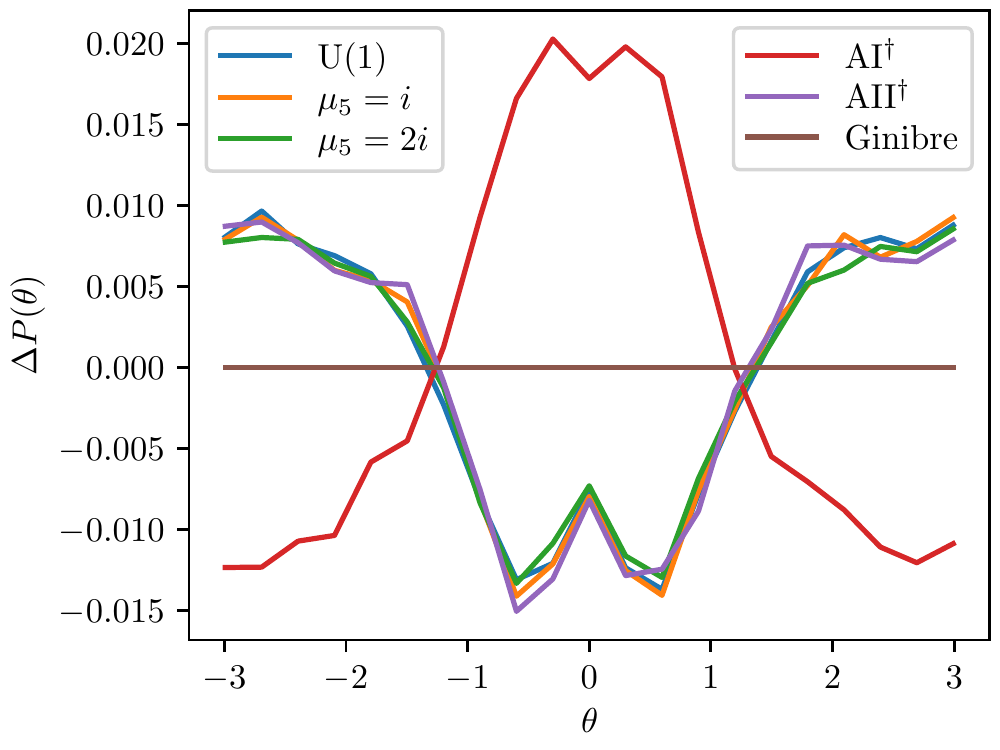}
    \end{tabular}}\hfill
  \parbox[c]{.4\textwidth}{\includegraphics[width=.4\textwidth]{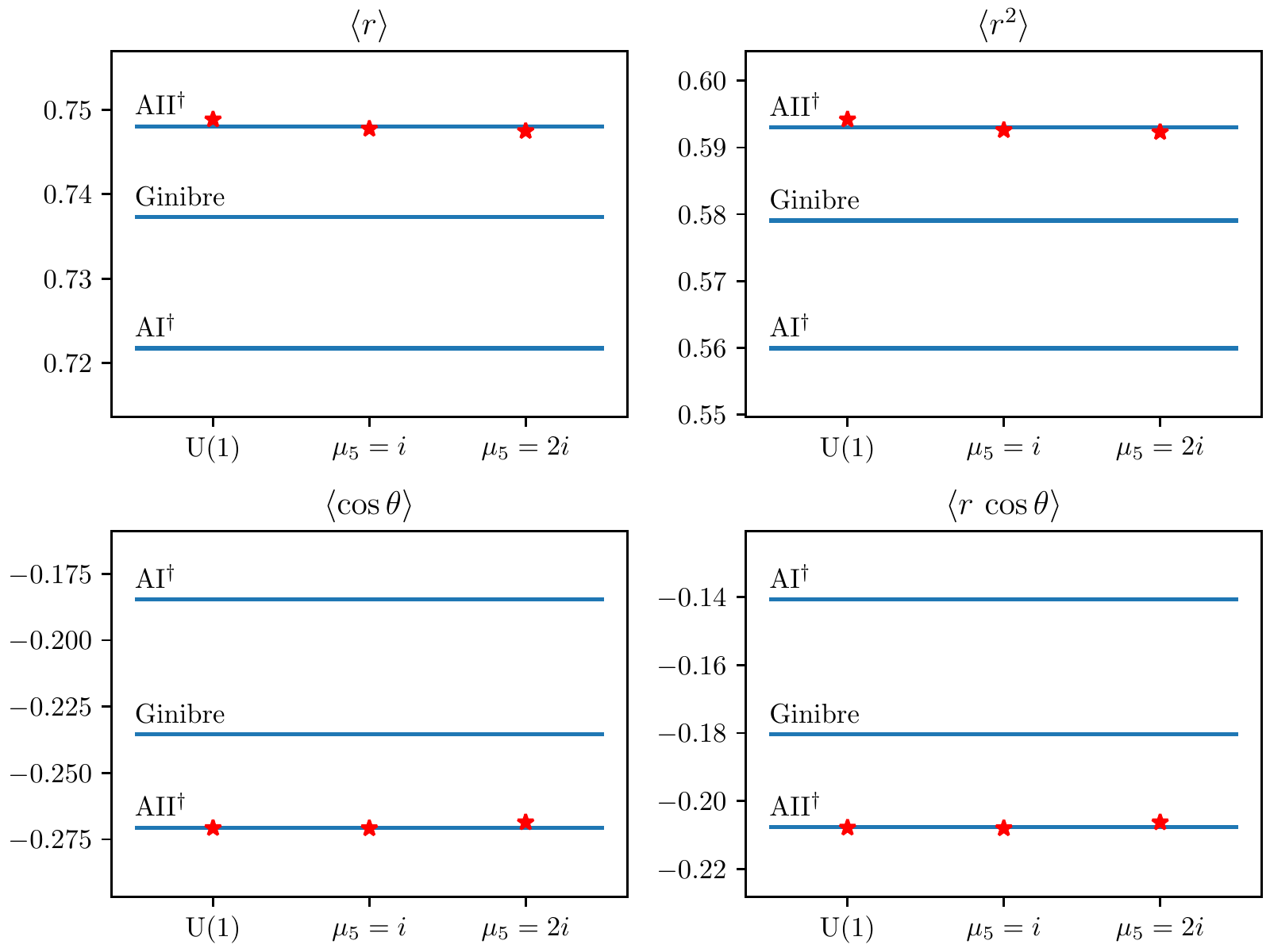}}
  \caption{Lattice results for SU(2) fundamental and our three cases (chiral U(1) field, $\mu_5=i$ and $\mu_5=2i$) for the marginal distributions $P(r)$ (first column) and $P(\theta)$ (second column) as well as for moments of $P(z)$ (third and fourth column), compared with the RMT predictions. The plots for $\Delta P(r)$ and $\Delta P(\theta)$ show differences with respect to the Ginibre class, which makes it easier to see that the lattice results agree with \AIId.}
  \label{fig:1dfund}
\end{figure}

\begin{figure}
  \centering
  \renewcommand\ig[1]{\includegraphics[width=.28\textwidth,valign=c]{#1}}
  \parbox[c]{.595\textwidth}{
    \begin{tabular}{c@{\hspace*{.2em}}c}
      \ig{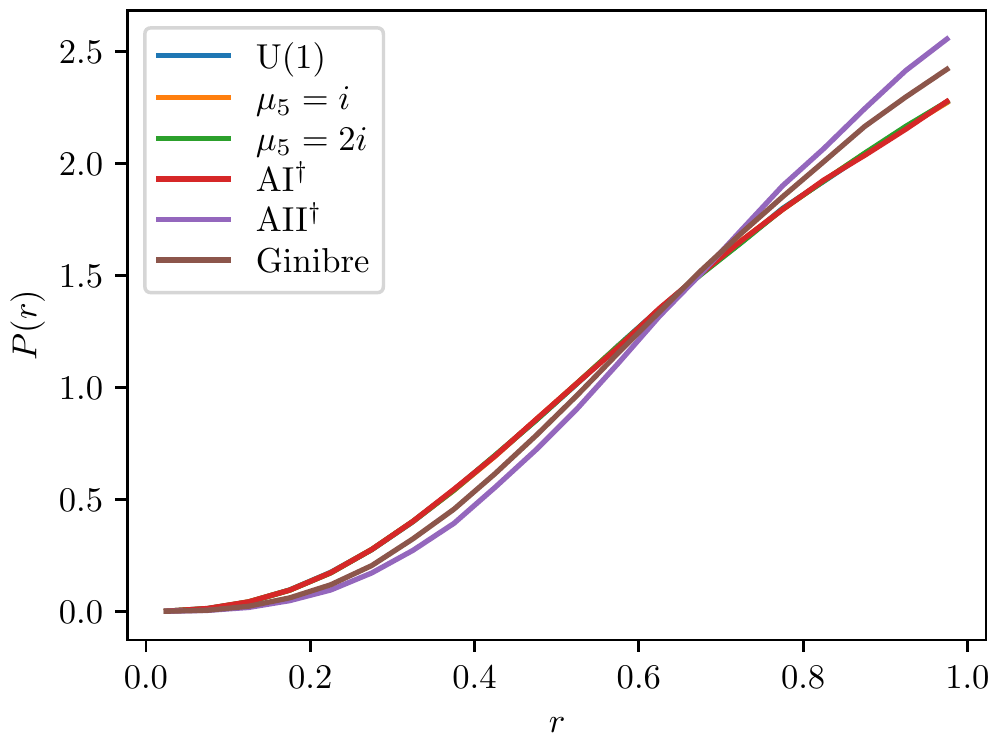} & \ig{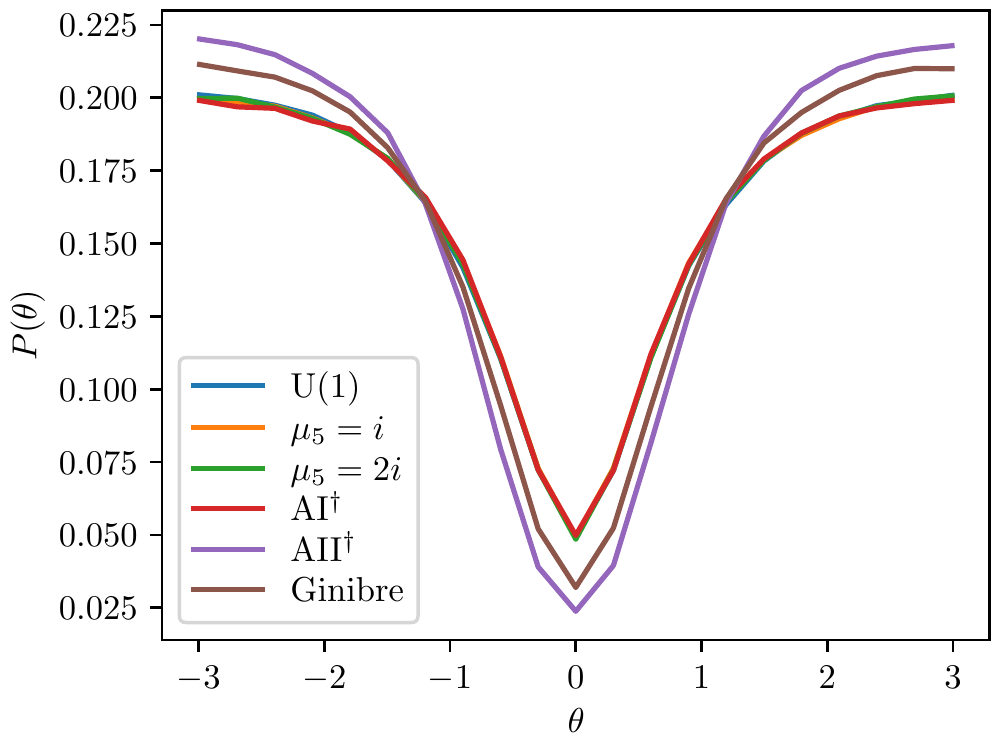} \\
      \ig{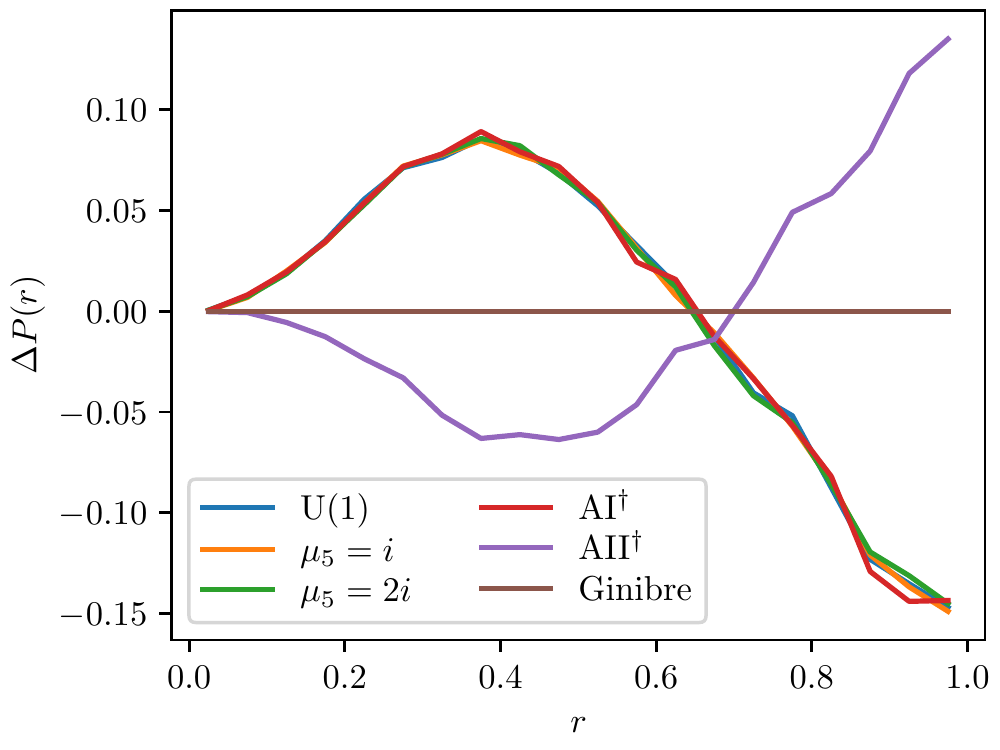} & \ig{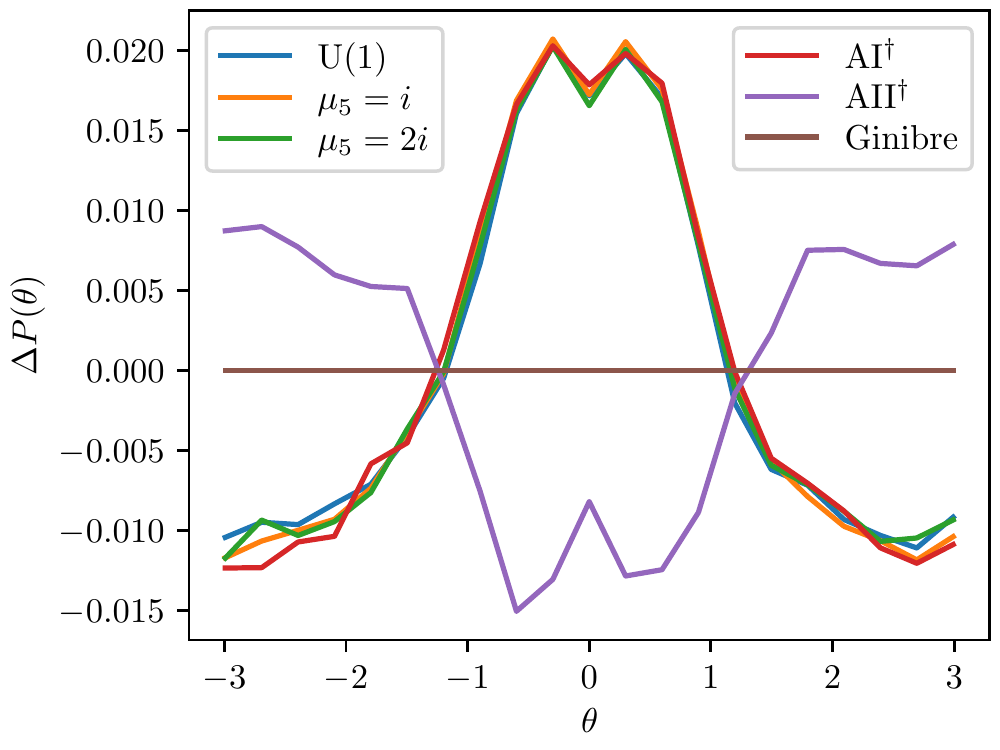}
    \end{tabular}}\hfill
  \parbox[c]{.4\textwidth}{\includegraphics[width=.4\textwidth]{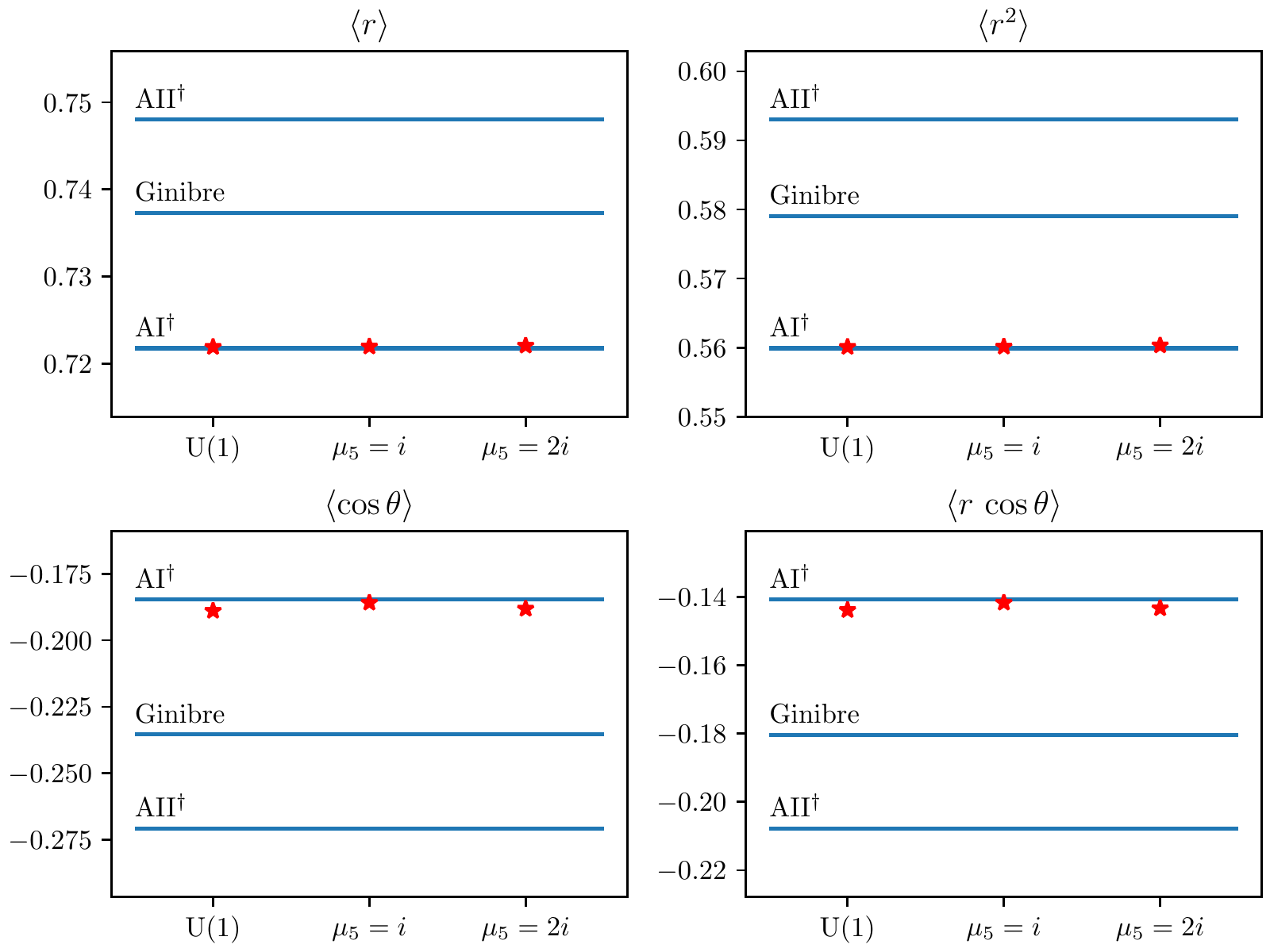}}
  \caption{Same as Fig.~\eqref{fig:1dfund} but for SU(2) adjoint. The lattice results now agree with universality class \AId.}
  \label{fig:1dadj}
\end{figure}

In Figure~\ref{fig:2d} we show the distribution $P(z)$ of the complex spacing ratio $z$ defined in Eq.~\eqref{eq:nn}. The plots confirm our prediction that for the staggered Dirac operator with chiral U(1) field or chiral chemical potential, the fundamental and adjoint representation of gauge group SU(2) corresponds to universality class \AIId and \AId, respectively. Since such two-dimensional plots are not completely unambiguous we also plot the marginal distributions $P(r)$ and $P(\theta)$ as well as several moments of $P(z)$, see Fig.~\ref{fig:1dfund} for SU(2) fundamental and Fig.~\ref{fig:1dadj} for SU(2) adjoint. These plots confirm our predictions within the numerical errors, which are partly statistical and partly due to the finite lattice, which leads to boundary effects in Eq.~\eqref{eq:nn} \cite{SRP2020}.

\section{Summary}

We have shown that the nonstandard universality classes \AId and \AIId of non-Hermitian RMT are realized in the bulk spectral correlations of the Dirac operator coupled to a chiral U(1) gauge field or an imaginary chiral chemical potential. In the 
continuum we find \AId for pseudoreal representations and \AIId for real ones. For the staggered lattice Dirac operator these symmetries are reversed. The numerical results of our lattice simulations for the complex spacing ratios of Eq.~\eqref{eq:nn} confirm our predictions for the universality classes. We have also derived novel spectral sum rules that serve as a useful check on the eigenvalues computed numerically.

In future work it would be interesting (a) to consider the deconfined phase, (b) to take a closer look at the spectral correlations near zero, and (c) to study the continuum limit, in which the ``correct'' universality classes should be recovered. It would also be interesting to derive analytical RMT results for $P(z)$.

\bibliographystyle{JHEP_lat21}
\bibliography{references}

\end{document}